\documentclass[useAMS,usenatbib]{mn2e}
\usepackage{graphicx}

\title[CCD Photometry of NGC 5466]{CCD Photometry of
 the globular cluster NGC~5466. RR Lyrae light curve decomposition and the distance scale\thanks{Based on
  observations collected at the Indian
   Astrophysical Observatory, Hanle, India.}}
\author[A. Arellano Ferro et al.]{A. Arellano Ferro$^{1}$\thanks{E-mail:
armando@astroscu.unam.mx}, V. Rojas L\'opez$^{1}$\thanks{E-mail:
victoria@astroscu.unam.mx}, Sunetra Giridhar$^{2}$, D.M. Bramich$^{3}$\\
$^{1}$Instituto de Astronom\1a, Universidad Nacional Aut\'onoma de M\'exico\\
$^{2}$Indian Institute of Astrophysics, Koramangala 560034, Bangalore, India\\
$^{3}$Isaac Newton Group of Telescopes, Apartado de Correos 321, E-38700 Santa Cruz de la Palma, Canary Islands, Spain \\}

\begin{document}
 
\date{Accepted . Received ; in original form }

\pagerange{\pageref{firstpage}--\pageref{lastpage}} \pubyear{2002}

\maketitle 

\label{firstpage}

\begin{abstract}

We report the results of CCD $V$ and $r$ photometry of the globular cluster NGC 5466.
  The difference image analysis technique adopted in this work has 
  resulted in accurate time series photometry even in crowded regions
  of the cluster enabling us to discover five probably semi-regular variables.
We present new photometry of three previously known eclipsing binaries and six SX Phe stars.
The light curves of the RR Lyrae stars have been decomposed in their Fourier harmonics and their fundamental physical parameters have been estimated using semi-empirical calibrations. The zero points of the metallicity, luminosity and temperature scales are discussed and our Fourier results are transformed accordingly.
The average iron abundance and distance to the Sun derived from individual RR Lyrae stars, indicate values of [Fe/H]=$-1.91 \pm 0.19$ and  
D = $16.0 \pm 0.6$ kpc, or a true distance modulus of $16.02 \pm 0.09$ mag, for the parent cluster. These values are respectively in the Zinn \& West  metallicity scale and in agreement with recent luminosity determinations for the RR Lyrae stars in the LMC.

The $M_V$-[Fe/H] relation has been recalibrated as $M_V=+(0.18\pm0.03)[Fe/H]+(0.85\pm0.05)$ using the mean values derived by the Fourier technique on RR Lyrae stars in a family of clusters. This equation predicts $M_V=0.58$ mag for [Fe/H]=$-1.5$, in agreement with the average absolute magnitude of RR Lyrae stars calculated from several independent methods.
The $M_V$-[Fe/H] relationship and the value of [Fe/H] have implications on the age of the globular clusters when determined from the magnitude difference between the horizontal branch and the turn off point (HB-TO method). The above results however would not imply a change in the
age of NGC~5466, of 12.5$\pm$0.9 Gyr, estimated from recent isochrone fitting.

\end{abstract}
      

\section{Introduction}

Determination of the age, distance and metallicity of globular clusters has enormous relevance in several astrophysical problems such as the construction of a model of the formation of the Galaxy, and the stellar structure and evolution on the Horizontal Branch (HB). The distribution of globular cluster ages as a function of galactocentric distance, the dispersion of cluster ages at a given metallicity and the role of a "second-parameter" (other than metallicity) in the structure of the HB, are aspects that may be constrained by our knowledge of these fundamental quantities.

It is known that the HB morphology in globular clusters of similar metallicities varies, and the identification of a second parameter responsible of this has originated much discussion in the astronomical literature. Recent findings that the more massive clusters tend to 
have HBs more extended to higher temperatures led Recio-Blanco et al. (2006) to suggest the cluster total mass may be  the second parameter. 
Considering the fact that the HB morphology varies
 due to stellar evolution, the age is still a strong candidate for the
 second parameter status
(e.g. Stetson et al. 1999; Catelan 2000 and references therein). 

Two populations of galactic globular clusters have been identified on the basis of their metallicities, kinematics and horizontal-branch structure, which has implications on the history of the formation of our Galaxy (see e.g. Zinn, 1993; van den Bergh 1993; Zinn 1996).
The determination of accurate absolute ages of globular clusters is, however, linked to our knowledge of the cluster distance scale (Gratton et al. 1997; Gratton 1998; Carretta et al. 2000; VandenBerg 2000; Gratton et al. 2003). Nevertheless, calculations of relative ages, i.e. cluster age differences relative to well studied nearby clusters (e.g. M92 and M5), at a fixed metallicity, provide an age scale  with an uncertainty $\leq$ 1 Gyr, which allows one to conclude that the outer-halo globular clusters, with galactocentric distances 
$R_G \geq$ 50 kpc, are 1.5-2 Gyr younger than the inner-halo clusters, $R_G \leq$ 10 kpc,  of the same metallicity (see Stetson et al. 1999; VandenBerg 2000 and references therein). As for the relative ages and their connection with globular cluster metallicities, it has been found that all clusters with [Fe/H] $\leq -1.7$ are old and coeval and on average 1.5 Gyr older than intermediate-metallicity clusters ($-1.7 \leq [Fe/H] \leq -0.8$).
All the clusters with [Fe/H] $\geq -0.8$ are ~1 Gyr younger than the most metal-poor ones. The age dispersions in each metallicity group are $\leq$ 1 Gyr (De Angeli et al. 2005). There is no correlation of the cluster age nor of metallicity with the galactocentric distance.

Direct tests to calibrate the cluster chronology should involve however, direct measurements of the metallicities of individual stars in the clusters and direct estimates of the distances and ages
of the systems. These efforts have been made recently, based on detailed spectroscopic analysis 
(James et al. 2004; Gratton et al. 2005; Yong et al. 2005; Cohen \& Melendez, 2005) and  deep photometric measurements of the main sequence and subgiant branch in clusters of different  metallicities and galactocentric distances (e.g. Stetson et al. 1999).

An alternative and efficient method to estimate these relevant physical parameters takes advantage of the fact that the light curve morphology of RR Lyrae stars is connected with fundamental stellar physical parameters, as first suggested by Walraven (1953).
The mass, effective temperature, luminosity, helium fraction and iron content
for RR Lyrae stars can be calculated from the parameters of the Fourier decomposition of their light curves using the semiempirical calibrations calculated in recent years (e.g. Simon \& Clement 1993, Jurcsik \& Kov\'acs 1996, Morgan et al. 2007). The present capabilities of CCD photometry and difference image analysis (Alard \& Lupton 1998,
Alard 2000, Bramich et al. 2005, Bramich 2007) enables us to perform accurate photometry of very faint stars in remote systems, even in the crowded central regions of globular clusters. There are now numerous works in which the physical parameters of RR Lyrae stars have been estimated from their Fourier light curve decomposition and consequently the mean distances and metallicities of the parent clusters have been determined (for a summary of these works see L\'azaro et al. 2006).

In the present work we report standard $V$ and instrumental $r$ CCD photometry of NGC 5466, perform the Fourier light curve decomposition of known RR Lyrae stars and report the discovery of new variable stars in the cluster. The globular cluster NGC 5466 (R.A.(2000)$=14^h05^m27^s$.3, DEC(2000)$=+28^{\circ} 32'04''$) is located in the intermediate galactic halo (l=42.15, b=+73.59, $Z=$ 15.9 kpc, R$_G$=16.2 kpc). It is a rather diffuse system that, up until 1918, was not considered a globular cluster but more likely an open cluster (Shapley 1918). Soon after, however, it was listed for the first time as "certainly a globular cluster" (Shapley 1919) and then described as a loose globular cluster (Shapley 1930). Its concentration class is XII and its spectral type F5 (Buonanno et al. 1984). From the periods of its RR Lyrae stars the cluster is of the Oosterhoff type II (Oosterhoff 1939). Its diffuse nature is most likely due to the effects of Galactic tides, an idea supported by the recent discovery of tidal tails around
the cluster (Belokurov et al. 2006; Grillmair \& Johnson 2006).

In the catalogue of variable stars in globular clusters (Clement et al. 2001; 2002) there are reported 32 variables in NGC 5466; 13 RRab, 6 RRc; 2 possible second overtone pulsators or RR2, 1 possible double mode pulsator or RRd,
1 anomalous cepheid; 3 eclipsing binaries and 6 SX Phe stars. 

Photoelectric photometry of the cluster dates back to the work of Cuffey (1961) in the $PV$ system. Buonanno et al. (1984) performed $BV$ photographic photometry aimed at producing a colour magnitude diagram and to measure accurate positions of the stars in the cluster. CCD photometry has been reported by Corwin et al. (1999). These authors determine the ephemerides of 20 RR Lyrae stars. In the present paper, we aim to revisit the light curves of known RR
Lyrae stars using the relatively new difference imaging technique and to Fourier decompose them into their harmonics, so as to
estimate individual physical parameters of astrophysical importance. We also search for new variables.

In Sect. 2 we describe the observations and data reductions. In Sect. 3 individual objects are discussed and new variables are presented. In Sect. 4 
we calculate the physical parameters using the Fourier light curve decomposition method and discuss the transformations to homogeneous scales. In Sect. 5 we
discuss the results in the wider context of the RR Lyrae distance scale and the age of the cluster. In Sect. 6 we summarize our conclusions.

%############################  tabla 1
\begin{table*} 
\footnotesize{
\begin{center}
\caption[{\small Standard $V$ and mean instrumental magnitudes $v$ per observing run for 11 standard stars taken from Buonanno et al. (1984). {\it n} is the number of images considered in the average.}] 
{\small Standard $V$ and mean instrumental magnitudes $v$ per observing run for 11 standard stars taken from Buonanno et al. (1984). {\it n} is the number of images considered in calculating the mean and standard deviation.}
\label{tab_estandar}
\hspace{0.01cm}
 \begin{tabular}{|lc|ccc|ccc|ccc|ccc|}
\hline
  & & $v$ & & & $v$ & & & $v$ & & &$v$& &  \\
Star$^*$ & $V$&H04& $\sigma_{\rm H04}$ & n &H05& $\sigma_{\rm H05}$ &n&K05& $\sigma_{\rm K05}$ & n &SPM05& $\sigma_{\rm SPM05}$& n \\
\hline
     F,21 &   14.11 & 15.012 &  0.003& 24 & 15.104 & 0.007& 36 &        &     &   &16.618 &  0.001&4 \\
     H,3 &   16.66 & 17.677 &  0.008 &41& 17.731 & 0.016& 50 &        &       & &       & &\\
     I,7 &   15.78 & 16.806 &  0.007& 42& 16.895 & 0.007& 48 &        &      &  &18.457  & 0.010&5\\
     J,11 &   16.87 & 17.879 &  0.011 &44& 17.974 & 0.022& 52 &        &       & &19.414  &  0.014&5\\
     M,28 &   17.64 & 18.641 &  0.038& 41& 18.767 & 0.027& 53 & 19.742 &  0.015&3 & 20.278 & 0.008&5\\
     O,231 &   14.28 & 15.255 &  0.004 &44& 15.385 & 0.006& 51 & 16.617 &  0.008&2 &16.866  & 0.004&5\\
     P,259 &   14.13 & 15.039 &  0.005& 43& 15.152 & 0.010& 51 & 16.585 &  0.010&3&16.719 &  0.005&5 \\
     Q,263 &   16.37 & 17.375 &  0.008 &44& 17.480 & 0.017& 53 & 18.667 &  0.007&3 &19.028 &  0.004 &5\\
     R,266 &   16.67 & 17.702 &  0.039& 43& 17.836 & 0.019&53 & 18.905 &  0.031&3 &19.271 &  0.007 &5\\
     S,F285 &   14.88 & 15.904 &  0.013 &43& 15.999 & 0.011&53 &        &       & &       & &\\
     Z,299 &   17.57 & 18.506 &  0.013& 43& 18.635 & 0.022 &51 &        &       & & 20.128 &  0.010 &5\\
\hline
\end{tabular}

\end{center}
* Star identifications with capital letters are from Cuffey (1961) while numbers are the equivalents in Buonanno et al. (1984).
}
\end{table*}

\section{Observations and Reductions}
\label{sec:Observations}

The observations employed in the present work were performed using the Johnson $V$ and $R$ filters and
obtained on three different telescopes. On April 8 and 9, 2004 (H04) and on May 14 and 15, 2005 (H05),
we used the 2.0m telescope of the Indian Astronomical Observatory (IAO), Hanle, India, located at 4500m above sea level. The estimated seeing was $\sim$1 arcsec.
The detector was a 
Thompson CCD of 2048 $\times$ 2048 pixels with a pixel
scale of 0.17 arcsec/pix and a field of view 
of approximately $11. \times 11.$ arcmin. On February 6, 7 and 8, 2005 (K05), we used the 1.0 m telescope of the the Vainu Bappu Observatory (VBO), Kavalur, India, located a 700m above sea level. The detector was a 
Thompson CCD of 1024 $\times$ 1024 pixels with a pixel scale of 0.37 arcsec/pix and a field of view of approximately $6.5 \times 6.5$ arcmin. The mean
seeing was $\sim$2.8 arcsec. On January 29 and 30, 2005 (SPM05) we used the 0.84m telescope of the San Pedro M\'artir Observatory (VBO), Baja California, Mexico, located a 2790 meters above sea level. The detector was a 
Thompson CCD of 2048 $\times$ 2048 pixels with a pixel scale of 0.39 arcsec/pix and a field of view of approximately $8 \times 8$ arcmin.
The estimated seeing was $\sim$2.0 arcsec but some telescope tracking problems produced elongated stellar profiles and
consequently we rescued only a handful of images.

\begin{figure} 
\includegraphics[width=8.cm,height=10.cm]{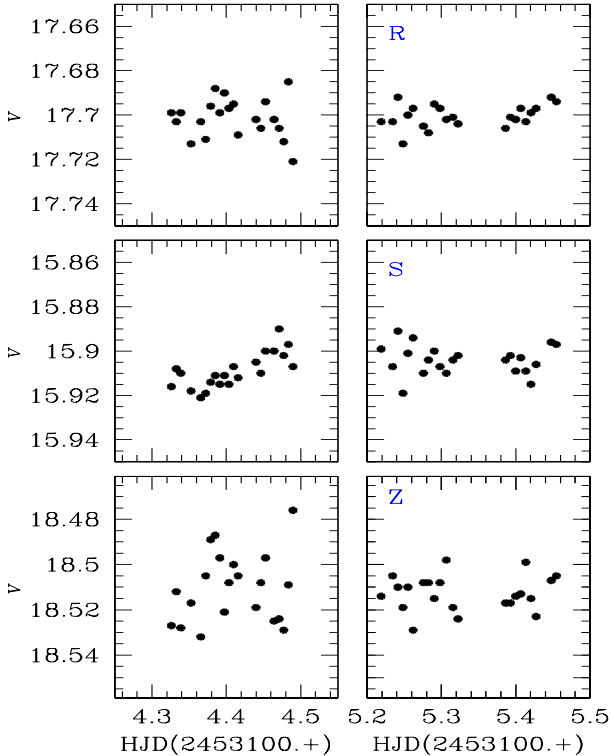}
%\vspace{4cm}
\caption{Instrumental magnitudes as a function of HJD for three standard stars on two consecutive nights from the H04 observing run.}
    \label{estand1}
\end{figure}

Differential photometry using the difference imaging technique (Alard \& Lupton 1998; Alard 2000), and
employing the methodology and pipeline used in Bond et al. (2001) and
Bramich et al. (2005), was performed.
This procedure involves
the matching of a high quality reference image to each image in the time
series, by solving for a spatially-varying convolution kernel and
differential sky background function. In these reductions we used a
kernel decomposed onto basis functions formed as a product of a
two-dimensional Gaussian function of kernel coordinates $u$ and $v$ with
a polynomial in $u$ and $v$. We used three Gaussian components with
$\sigma$ values of 2.1, 1.3 and 0.7 pixels and associated polynomial degrees of 2, 4 and 6, respectively.
We modeled the differential sky background as a polynomial of degree 2 in spatial
coordinates $x$ and $y$. To allow for the kernel's spatial dependence, the
coefficients of the kernel basis functions were polynomials of degree 2 in $x$ and $y$.

Difference images were constructed
via the subtraction of the convolved reference image from the time series
images. Photometry on the difference images via optimal PSF scaling (Bramich et al. 2005) yielded
light curves of differential fluxes for each star. Conversion of the
light curves to the magnitude scale requires an accurate measurement of the
reference flux. This was achieved as in Bramich et al. (2005) by using the PSF fitting package
{\tt DAOPHOT} (Stetson 1987). The corresponding
equations we used to convert from differential fluxes to magnitudes are
as follows:

\begin{equation}
f_{\mbox{\small tot}}(t) = f_{\mbox{\small ref}} + \frac{f_{\mbox{\small diff}}(t)}{p(t)},
\end{equation}

\begin{equation}
m(t) = 25.0 - 2.5 \log (f_{\mbox{\small tot}}(t)),
\end{equation}

\noindent
where $f_{\mbox{\small tot}}(t)$ is the star flux (ADU/s) at time $t$, $f_{\mbox{\small ref}}$ is the star flux (ADU/s)
as measured on the reference image, $f_{\mbox{\small diff}}(t)$ is the differential flux (ADU/s) at time $t$ as
measured on the difference image, $p(t)$ is the photometric scale factor (the integral of the kernel
solution over $u$ and $v$) at time $t$ and $m(t)$ is the magnitude of the star at time $t$.
Uncertainties were propagated in the correct analytical fashion.

% TABLA 2
\begin{table}
\footnotesize{
\begin{center}
\caption[{\small }] {\small Ephemerides of the known RR Lyrae stars in NGC 5466. Some periods have been revised in this work (column 4) and new epochs were calculated for all the stars (column 5).}
\label{new_epoch}
\hspace{0.01cm}
 \begin{tabular}{llccc}
\hline
 ID & Period & Epoch  &  New Period & New epoch \\
& (days) & (+240~0000) & (days) & (+240~0000) \\
\hline
V2 & 0.588502$^1$ & 40684.221 &    & 40683.793 \\
V3 & 0.5780645$^2$ & 40704.319 & & 40704.855 \\
V4 & 0.5113067$^1$ & 40704.461 & & 40704.597 \\
V5 & 0.6152241$^2$ & 39945.659 & & 39945.992 \\
V6 & 0.6209516$^2$ & 40705.408 & & 40705.876 \\
V7 & 0.7034205$^1$ & 40702.398 & & 40702.402 \\
V8 & 0.6291182$^1$ & 40705.358 & & 40705.379 \\
V9 & 0.6850366$^2$ & 39947.328 & & 39947.996 \\
V10& 0.7092782$^2$ & 40705.468 & & 40705.504 \\
V11& 0.60771$^2$ & 40705.285 & & 40705.871  \\
V12& 0.2942387$^1$ & 39945.210 &0.293378 & 39945.256\\
V13 &0.3415568$^2$ & 40736.379 & & 53087.421  \\
V14 &0.7858557$^2$ & 39947.568 & & 48382.894  \\
V15 &0.398625$^2$  & 40705.223 & 0.40185& 40705.491 \\
V16 &0.424530$^2$  & 39945.372 & 0.419999 & 39946.137 \\
V17 &0.3701062$^2$ & 40706.394 & & 53085.994 \\
V18 &0.3744435$^2$ & 30519.697 & 0.37270 & 53085.954\\
V19 &0.8213010$^3$ & 40705.737 &  & 40706.113  \\
V20 &0.229496$^2$  &           & 0.30208 & 49876.976 \\
V21 &0.791027$^2$  &           &  & 48382.785  \\
\hline
\end{tabular}
\end{center}
Period sources:  1) Sawyer Hogg (1973), 2) Corwin et al. (1999), 3) McCarthy \& Nemec (1997).
}
\end{table}

The instrumental $r$ magnitudes were retained in the instrumental system since no absolute photometry was performed. No stars with R photometry in the field of the cluster were found in the literature.

\begin{figure*} 
\includegraphics[width=10.cm,height=10.cm]{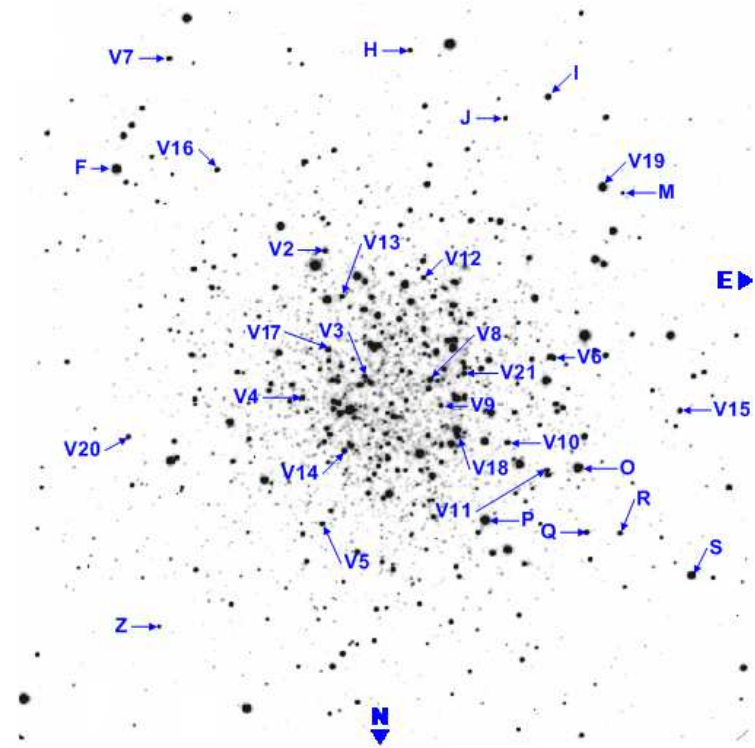}
\caption{The field of NGC 5466 with known RR Lyrae stars (names starting with $V$) and standard stars (capital letters) identified. The identifications of the standard stars is that of Cuffey (1961), but see also Table 1. The image was obtained at the 2.0-m HCT of the IAO.}
    \label{FIELD}
\end{figure*}

The instrumental $v$ magnitudes were converted to the Johnson $V$ standard system by 
using 11 selected standard stars from the list of Cuffey (1961) and in the field of our images of NGC 5466 in the H04 and H05 data. These stars have magnitudes in the interval $17.6 > V > 14.1$ and their $V$ standard magnitudes are listed by Buonanno et al. (1984). Unfortunately not all 11 stars were included in the smaller fields of view of the VBO and SPM images. The transformations to the standard system were performed for each observing run. Instrumental magnitudes of the standard stars on each image of a given run were plotted to check their constancy and dispersion. An example is shown in {Fig. ~\ref{estand1}} for the stars R, S and Z for the H04 observing run. The 11 stars considered are identified in {Fig. ~\ref{FIELD}}. {Table~\ref{tab_estandar}} lists the standard V magnitude taken from Buonanno et al. (1984), the average instrumental magnitude, the standard deviation of the mean and the number of images involved for each standard star for each observing run. The standard deviations give a good indication of the precision of the photometry.

The photometric transformation equation had a linear form
 $M_{std} = A~m_{ins} + B$, without the need of a quadratic 
term; moreover, the standard stars span a $B-V$ range from $0.04$ to $1.08$
and no significant colour term was found. In all cases the linear correlation coefficients were of the order of 0.999.

\begin{figure*} 
\includegraphics[width=16.cm,height=14.cm]{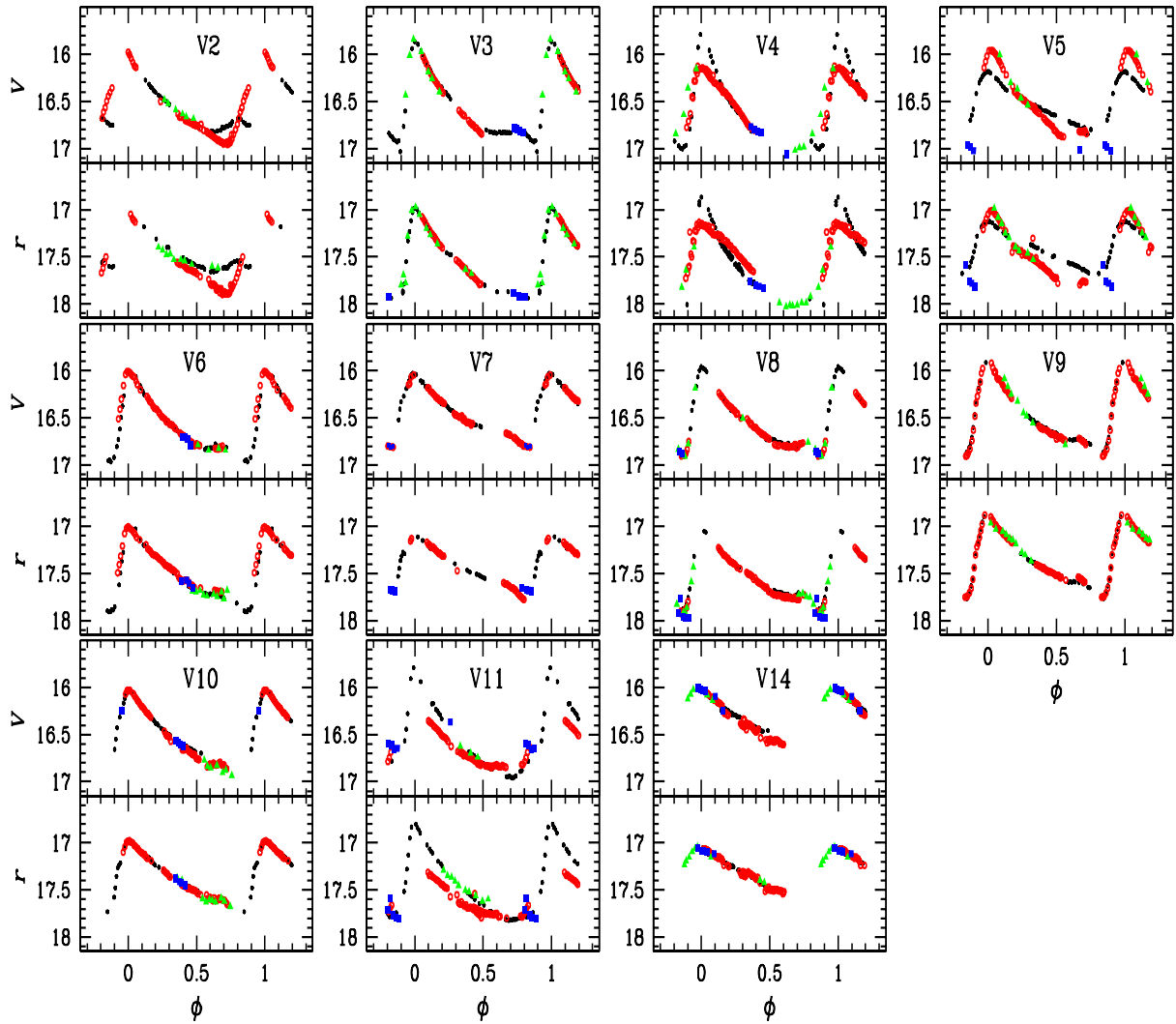}
%\vspace{4cm}
\caption{Standard $V$ and instrumental $r$ light curves of known RRab stars in NGC 5466. They have been phased 
with the new ephemerides in Table 1. The vertical scale is the same for all the stars. 
Dots correspond to the H04 season, circles to H05, triangles to K05 and squares to SPM05.}
    \label{var1}
\end{figure*}

\begin{figure*} 
\includegraphics[width=16.cm,height=10.cm]{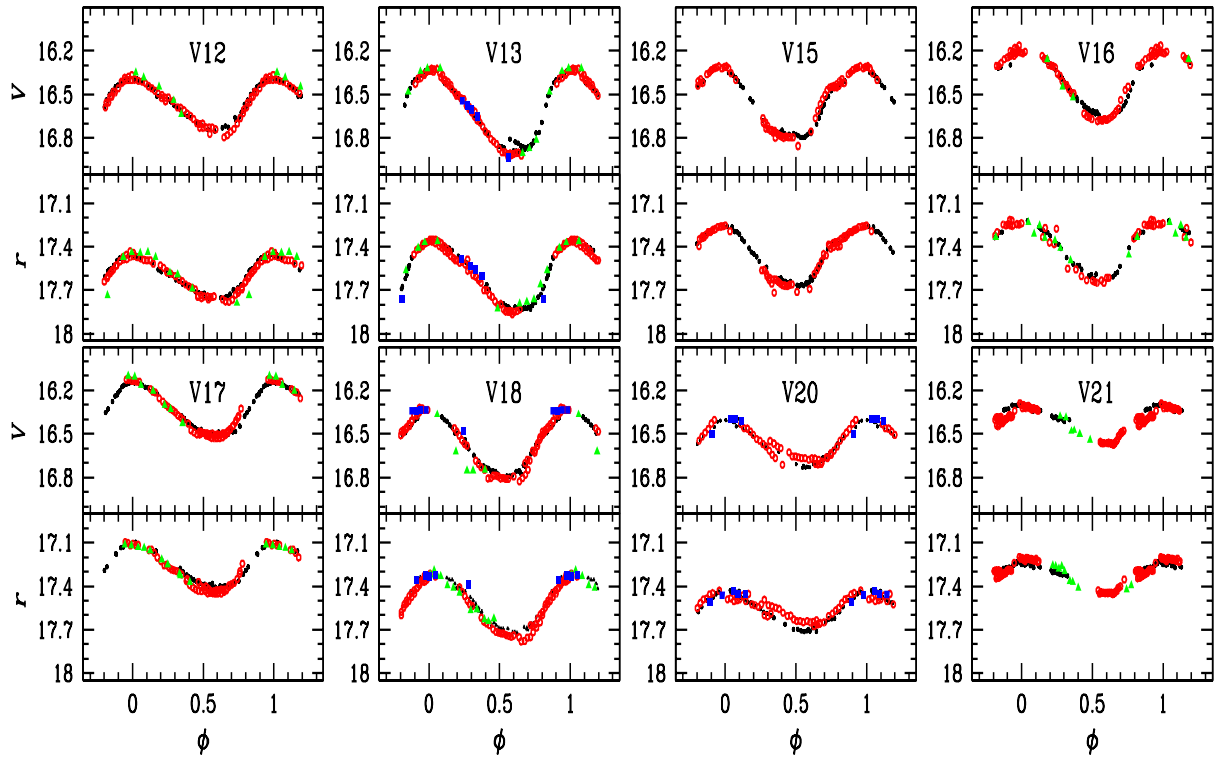}
\caption{Standard $V$ and instrumental $r$ light curves of previously known RRc stars in NGC 5466. 
Symbols are as in Fig. 3}
    \label{var2}
\end{figure*}

In {Figs.~\ref{var1}} and {\ref{var2}} the light curves in standard $V$ and instrumental $r$ magnitudes for the known RRab and RRc stars respectively, are displayed. 
Different symbols are used for the observations of each observatory as indicated in the captions.

In {Table ~\ref{new_epoch}} the ephemerides of the variable stars are reported. In a few cases the periods have been revised as will be discussed in the following section.

All our $V$ and $r$ photometry for the known RR Lyrae stars is available in Tables 8 and 9 respectively. The V photometry of the new and suspected variables is given in Table 10 and for the
known eclipsing binaries V28, V29, V30 and the SX Phoenicis stars it is given in Table 11. Tables 8 -- 11 are available in electronic format only and small portions of them are included at the end of this paper. 

\section{Discussion of individual objects and new variables.}

Variables V2, V4, V5 and V11 are known to exhibit the Blazhko effect (Sawyer Hogg 1973; Corwin et al. 1999). This is evident in their light curves in {Fig. ~\ref{var1}}. Except for V12, V15, V16, V18 and V20, the light curves of all the stars were properly phased with the previous ephemerides, indicated in {Table ~\ref{new_epoch}}. These five stars and V17 deserve some comments.

\subsection{V12} We used the data from Corwin et al. (1999) and our data from H04, H05 and K05 to determine a new period of 0.293378 $d$. The new ephemerides are used in {Fig. ~\ref{var2}}. Small variations in amplitude are hinted at.

\subsection{V15} The period 0.398625 $d$ (Sawyer Hogg 1973) does not phase the present data correctly. We used data from H04 and H05 to determine the new period 0.40185 $d$.

\subsection{V16} This RRc star could not be phased properly using the reported period of 0.424530 $d$ (Corwin et al. 1999). Different phasing of the $V$ and $r$ light curves from different runs is observed when the old period is used. Using exclusively our data from H04, H05 and K05 we have estimated a new period of 0.419999 $\pm 10^{-6}$ d. The light curve is shown in {Fig.~\ref{var2}}

\subsection{V17} The period 0.3701062 $d$ (Corwin et al. 1999) correctly phases our data from seasons H04, H05 and K05. However, we point out that the amplitude in H05, 0.405 mag, is significantly larger than the amplitude derived from the H04 data, 0.366 mag ({Fig.~\ref{var2}}). This effect may be caused by double mode pulsation although we have not succeeded in finding two frequencies. Alternatively, the amplitude variation may indicate the Blazhko effect. Although the effect is not common among RRc stars, some cases have been reported (Moskalik \& Poretti 2003). More data would be required to settle the case. The star is also brighter by about 0.25 mag in the V filter than other RRc stars in the cluster. This might be an indication of evolution off the HB or contamination by a blended star. This was also noticed by Corwin et al. (1999) who considered the star as a blend on the basis of its colour, temperature and elongated appearance in their images. The star will not be considered in calculating the average values of the physical parameter in Table~\ref{fisicosC}.

\subsection{V18} The period 0.3744435 $d$ (Corwin et al. 1999) does not phase properly the present data. As for V16, using exclusively our data from H04, H05 and K05 we have revised the old period to find 0.37270 $\pm 10^{-5}$ $d$ which produces a more coherent light curve. The new period phases well Corwin et al's data but with a drift of about 0.15 in phase, which hints at a secular variation of the period.

\subsection{V19} 
\begin{figure} 
\begin{center}
\includegraphics[width=4.2cm,height=7.2cm]{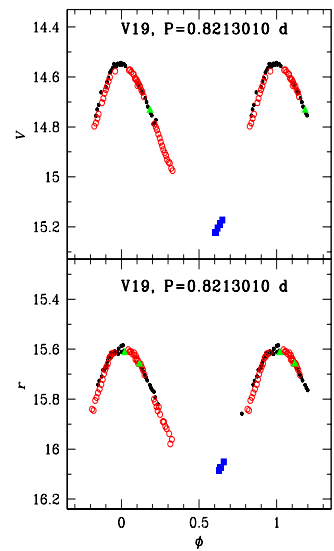}
\caption{$V$ and $r$ light curves of the anomalous cepheid V19. Symbols are as in Fig. 3.}
    \label{V19}
\end{center}
\end{figure}

This star is a known anomalous cepheid which has been the subject of a detailed abundance and period analysis (McCarthy \& Nemec 1997). In {Fig.~\ref{V19}} we present the light curve derived from our data phased with the period $0.8213010 \pm 0.0000003 d$ inferred by these authors.

\begin{figure} 
\includegraphics[width=8.cm,height=4.cm]{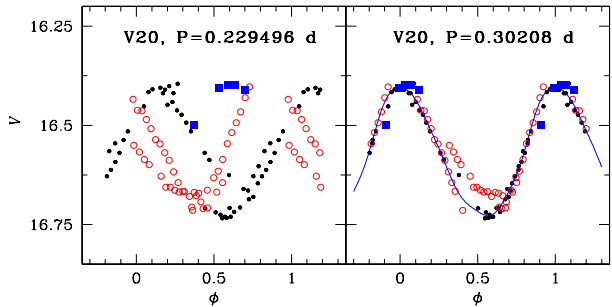}
\caption{Light curves of the peculiar star V20 phased with the old and new periods. Symbols are as in Fig. 3. See text for discussion.}
    \label{V20}
\end{figure}

\subsection{V20} This star was reported by Corwin et al. (1999) as peculiar since, in 1977,
they did not detect variation on two consecutive nights and noticed a sudden increase of brightness on the third night. They estimated a period of 0.229496 d, which produces a very poor light curve on Corwin et al's data and evidently does not phase our present data either. Instead, we have calculated a period of 0.30208 $\pm 10^{-5} d$, ({Fig.~\ref{V20}}). Despite a much better appearance of the light curve we note that two branches are produced even on a single observing run (H05). It is likely that the star is a member of a multiple system. Due to this peculiarity we have fitted the light curve only for the data from H04 in order to calculate the Fourier parameters, but the star will not be considered in calculating the average values of the physical parameters in Table~\ref{fisicosC}.

\subsection{Newly discovered variables}

We have detected clear light variations in 2 stars and possible variations in 3 stars in the field of NGC~5466, not
previously reported as variables. 
 In {Table~\ref{nuevasvar}}, we report the period, epoch
of maximum light, and the type of variability for each new variable.  The error on the
 periods of the new variables is about $\pm 4 \times 10^{-4}$ d.

\begin{figure} 
\includegraphics[width=8.5cm,height=8.5cm]{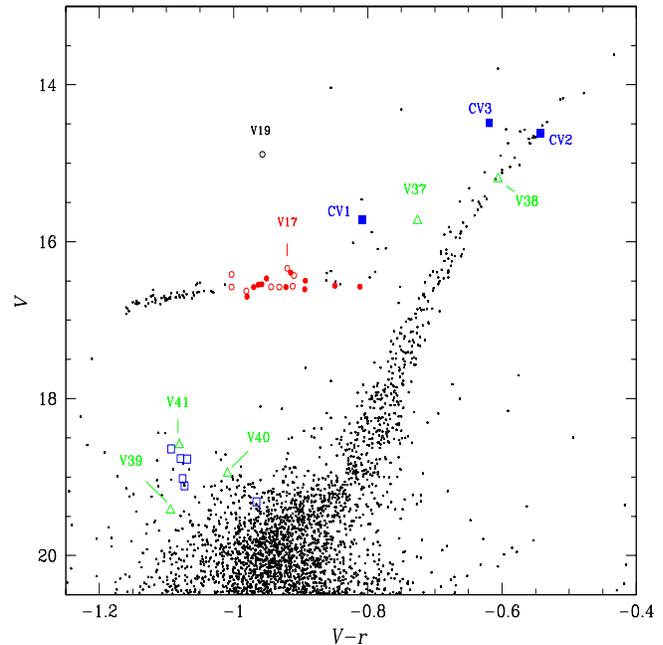}
\caption{($V-r$) vs $V$ colour magnitude diagram of NGC 5466. RRab stars (filled circles), RRc stars (open circles), new variables (open triangles), eclipsing variables (filled triangles), candidate variables (filled squares) and known SX Phe stars (empty squares) are shown. V17 is a peculiarly bright RRab. V19 is a known anomalous cepheid. See text for discussion.}
    \label{CMD}
\end{figure}

With the aim of identifying the type of variability we have built an $ad~hoc$ color-magnitude diagram (CMD). Since we have not reduced our $r$ photometry to the standard system we use the colour ($V-r$). The $V$ vs. ($V-r$) CMD is shown in {Fig.~\ref{CMD}}. The red giant, horizontal and asymptotic giant branches are clearly distinguished. Variable stars are shown and coded as indicated in the figure caption.

\begin{figure*}
\begin{center}
\includegraphics[width=11.cm,height=11.cm]{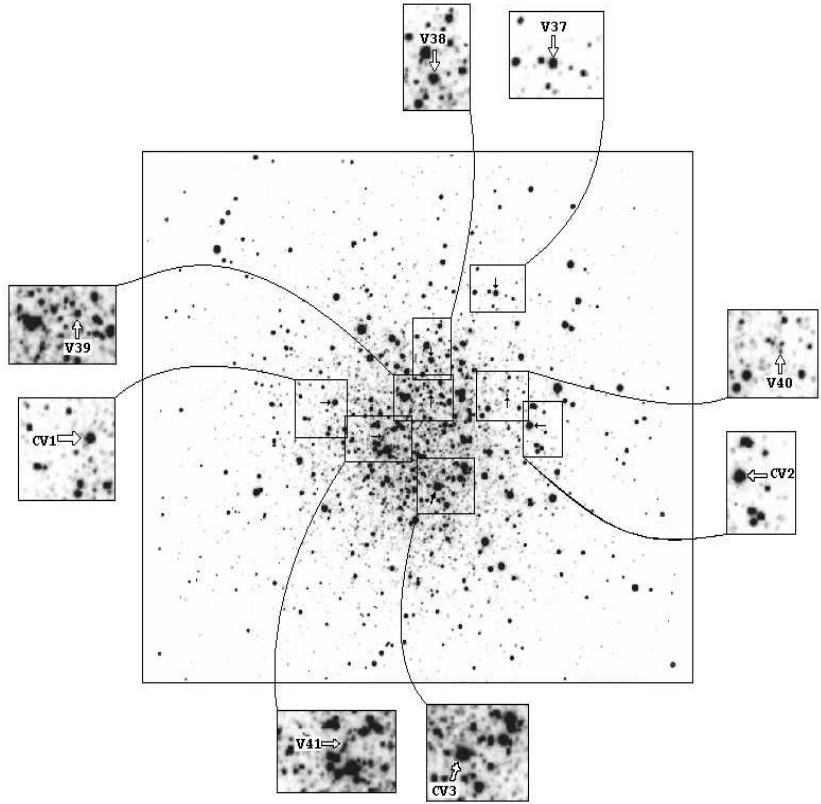}
\caption{Detailed identifications of new variables in NGC 5466. The location of the three known eclipsing vinaries V28, V29, and V30 is also indicated. The orientation of the field is like in {Fig.~\ref{FIELD}}.}
\label{nuevas}
\end{center}
\end{figure*}

The corresponding identifications of the new variables are shown in {Fig.~\ref{nuevas}}. The light curves are shown in {Fig. ~\ref{CURVENEWS}}. For the new variables we followed the identification system of Clement (2002). The three stars with possible variability are retained as candidates CV1, CV2 and CV3.
In the following paragraphs we offer some comments on each star.

\subsubsection{V37} In the subtracted images there are no significant residuals that might indicate a poor subtraction. The period analysis reveals two clear periods; P$_0=$ 0.3224 $d$ and P$_1=$ 0.2398 $d$, which happen to have a period ratio   $P_1/P_0=$ 0.744. The light curve is shown in Fig. ~\ref{CURVENEWS}. The star is ~0.75 mag brighter than the RR Lyrae stars in the cluster.  The possibility that it is a foreground double mode RR Lyrae star or RRd can be discarded by its  fundamental period being too short, which would place the star outside the Petersen diagram (Petersen 1973) for known stars. Also its colour ($V-r$) being too red. The alternative possibility that it is a foreground double mode $\delta$~Scuti can probably be disregarded since the star is too  red and it would not fall on the instability strip .
Its position on the CMD suggests instead that it is a red semiregular variable.

\subsubsection{V38} This star has a period of 0.3777 $d$.  Its light curve is shown in Fig. ~\ref{CURVENEWS}. 
Despite of being a brighter object its light curve has large scatter hinting possibly to the presence of more than one periodicity. From its position on the CMD the star could be a red semiregular variable.

\subsubsection{CV1, CV2 and CV3} These three stars are 1 to 2 mag brighter that the RR Lyrae stars in the cluster, 
hence our photometric accuracy is very high. Therefore notwithstanding the
 low amplitude of variation, we consider them as candidate variable stars.
 More observations of similar precision are required to firmly establish their variable nature.

\begin{figure} 
\includegraphics[width=8.cm,height=12.cm]{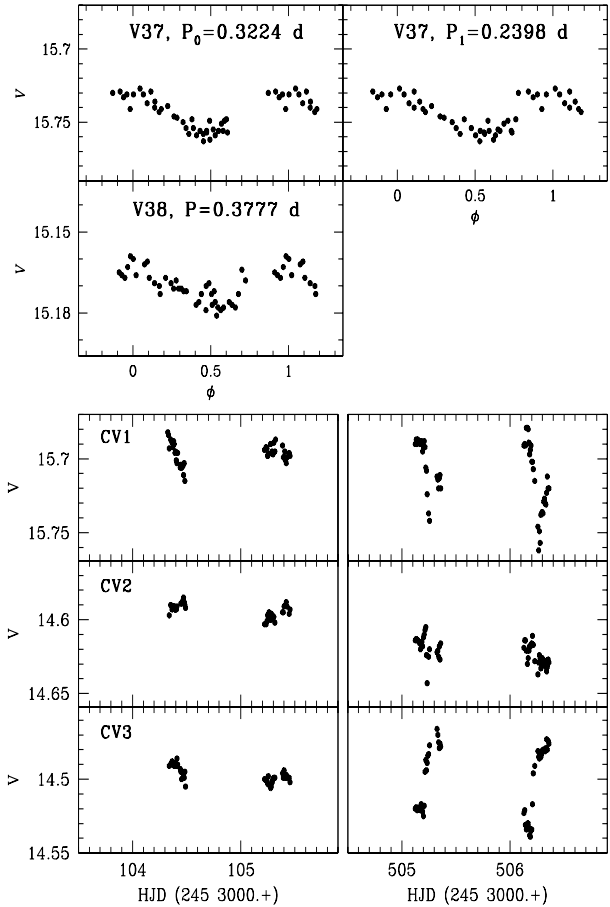}
\caption{$V$ light curves of the new and suspected variables from the H04 run. V37, probably a double-mode red variable, is phased with the two possible periods. V38 is probably a red semiregular variable. CV1, CV2 and CV3, are red semiregular variables. See text for details.}
    \label{CURVENEWS}
\end{figure}

In {Fig.~\ref{CURVENEWS}} we show the apparent light variations in the $V$ filter on the four nights from the H04 and H05 runs for each star. We are unable to find any periodicity in our data for these three stars. Their positions on the CMD suggest they may be red semiregular variables.

\begin{table}
\footnotesize{
\begin{center}
\caption[Nuevas variables] {\small Periods, epochs and classification of newly detected variables.}
\label{nuevasvar}
\hspace{0.01cm}
\begin{tabular}{llll}
\hline
Star & Period & Epoch & variable type\\
 & (days) & 2400000.0+  & \\
 & (days) & (days) & \\
\hline
V37 & 0.3224 & 53000.158 & red variable\\
    & 0.2398 & 53000.000 & \\
V38 & 0.3777 & 53000.035 & red variable\\
CV1 &  &  & red semirregular variable\\
CV2 & &  & red semirregular variable\\
CV3 &  &  & red semirregular variable\\
\hline
\end{tabular}
\end{center}
}
\end{table}

\subsection {Eclipsing Binaries}

\begin{figure} 
\includegraphics[width=8.cm,height=6.cm]{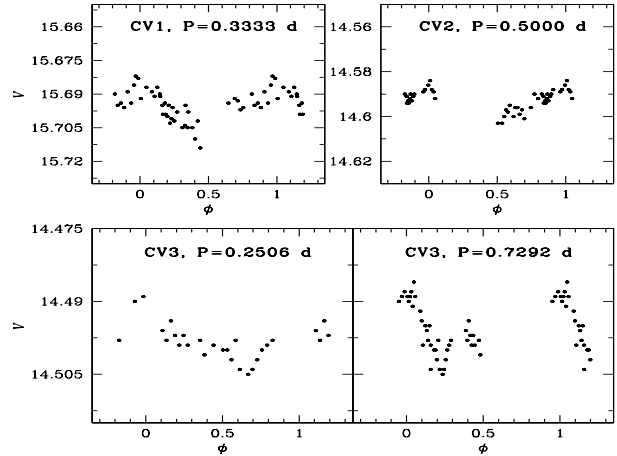}
\caption{H04 (filled circles) and H05 (open circles) $V$ data of the present paper for the three eclipsing binaries known in NGC 5466, phased with the periods and epochs given by Mateo et al. (1990). See text for details.}
    \label{binaries}
\end{figure}

The three eclipsing bianries V28, V29 and V30, for which we have retained the nomenclature of Clement et al. (2002), were discovered by Mateo et al. (1990) and characterized by Kallrath et al. (1992) (named respectivelly as NH19, NH30 and NH31).  V28 and V29 are two W Uma systems with periods
0.342144 $d$ and 0.297636 $d$ respectively. V30 is a short-period Algol system probably semidetached with a periodicity of 0.511320 $d$. In {Fig.~\ref{binaries}} we plot our H04 and H05 data with the above periods and with the epochs given by Mateo et al. (1990). We note that for V28 the primary eclipse is shifted by $\sim$0.1 and for V30 the shift is $\sim$0.2 in phase. These shifts are significant and most likely due to secular variations in the orbital periods which can be detected after the 18 years elapsed from the observations of Mateo et al.  and ours.

\subsection{SX Phoenicis stars}

\begin{figure} 
\includegraphics[width=8.cm,height=10.cm]{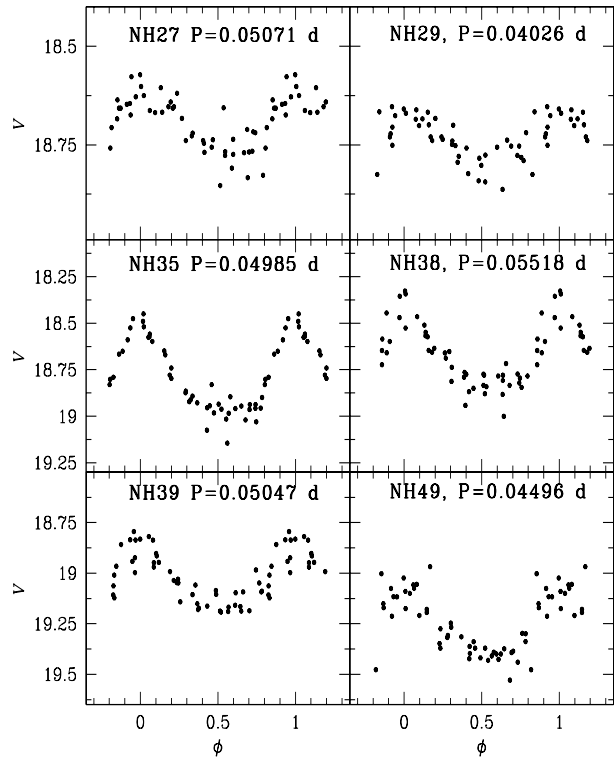}
\caption{$V$ light curves from the H04 run of the six SX Phe stars detected in our photometry. The periods are from Jeon et al. (2004).}
    \label{SXPHO}
\end{figure}

During the process of searching for new variables, we found six faint stars (V=18.5-19.3)
with short periods ($\sim$ 0.05 d). They turned out to be the SX Phoenicis stars reported by Jeon et al. (2004). We show in {Fig.~\ref{SXPHO}}  the light curves of the six stars detected in our photometry
(NH27, NH29, NH35, NH38, NH39 and NH49). According to these authors, the stars are multiperiodic. We have phased the light curves with the periods determined by Jeon et al. (2004). 
The mean magnitudes and the position of these stars in the blue straggler region (see Fig.~\ref{CMD}), agree very well with the measurements of Jeon and collaborators. The $V$ data of these stars is available in Table
11 (only in electronic form). The other 3 SX Phoenicis reported by Jeon et al. (2004) (SXP1, SXP2 and SXP3)
are all fainter than V=19 and hence we could not detect them in our data.

\section{Fourier light curves decomposition and physical parameters for the RR Lyrae stars}

 The mathematical representation of the light curves is of the  form:

\begin{equation}
m(t) = A_o ~+~ \sum_{k=1}^{N}{A_k ~cos~( {2\pi \over P}~k~(t-E) ~+~ \phi_k ) },
\end{equation}

\noindent
where $m(t)$ are magnitudes at time $t$, $P$ the period and $E$ the epoch. A linear
minimization routine is used to fit the data with the Fourier series model, deriving
the best fit values of $E$ and of the amplitudes $A_k$ and phases $\phi_k$ of the sinusoidal components. 

The fits are not shown in {Figs.~\ref{var1}} and {\ref{var2}} for simplicity and clarity.
From the amplitudes and phases of the harmonics in eq. 3, the Fourier parameters, 
defined as $\phi_{ij} = j\phi_{i} - i\phi_{j}$, and $R_{ij} = A_{i}/A_{j}$, 
were calculated. 
The mean magnitudes $A_0$, and the Fourier light curve fitting parameters of the individual RRab and
 RRc type stars in
$V$ are listed in {Table ~\ref{foufit}}. The mean dispersion about the fits is $\pm 0.020$ mag.
A comparison with the mean V magnitudes of Corwin et al. (1999) shows that our magnitudes are fainter by $0.043 \pm 0.044$ mag. The origin of this small difference is most likely due to the different approaches to the standarization of the instrumental magnitudes. Corwin and collaborators used 4 stars with photographic magnitudes taken from Buonanno et al. (1984).

\begin{table*}
\footnotesize{
\begin{center}
\caption[{\small Par\'ametros de los ajustes Fourier calculados para las curvas
de luz en el filtro V de las estrellas tipo {\bf RRc} del c\'umulo globular NGC
5466. Donde n es el n\'umero de arm\'onicos usados para el ajuste.}] {\small Fourier fit parameters for the V light curves.\label{foufit}}
\label{parametrosc}
\hspace{0.01cm}
 \begin{tabular}{cccccccc}
\hline
Star(n)&$A_0$ &$A_1$ & $A_4$ & $\phi_{21}$ & $\phi_{31}$& $\phi_{41}$& $D_m$ \\
\hline
&&&RRc stars&&&&\\
\hline
V12(8) & 16.577 & 0.176 & 0.003 & 4.56 & 2.77 & 2.08&\\
V13(4) & 16.629 & 0.283 & 0.016 & 4.58 & 2.38 & 1.82&\\
V15(4) & 16.568 & 0.246 & 0.020 & 4.12 & 4.13 & 2.94&\\
V16(8) & 16.417 & 0.240 & 0.003 & 5.30 & 3.91 & 2.47&\\
V17(8) & 16.340 & 0.185 & 0.004 & 4.17 & 1.29 & 4.67&\\
V18(8) & 16.578 & 0.236 & 0.007 & 4.38 & 2.90 & 3.40&\\
V20(8) & 16.576 & 0.162 & 0.002 & 3.94 & 2.36 & 5.45&\\
\hline
&&&RRab stars&&&&\\
\hline
V3(9)  & 16.579&  0.370 &0.102 & 3.68&    1.63 &  5.66&2.6\\
V6(9)  & 16.606 & 0.327  &0.070 &3.88 &   1.91  & 6.21&0.9  \\
V7(7)  & 16.469  &0.261 & 0.042 &4.10  &  2.39   &0.58&1.6   \\
V8(8)  & 16.547&  0.327& 0.076&  3.98   & 1.76   &6.03&2.0 \\
V9(9)  & 16.497 & 0.344 & 0.074& 3.93    &1.87&   0.03&2.8 \\
V10(7) & 16.573  &0.344 &0.050 & 4.28&    2.33 &  0.71&4.6 \\
V14(9) & 16.395&  0.277  & 0.043& 4.40 &   2.51  & 0.81&2.6 \\
\hline
&&&Blazhko stars&&&&\\
\hline
V2(8)   &16.542 &0.3737 & 0.0103 & 4.20 &2.42 & 5.82 &9.5\\
V4(7)   &16.666 &0.4433 & 0.1102 & 3.5515 &7.4723 & 5.1155 &2.7\\
V5(4)   &16.578 &0.3806 & 0.0363 & 3.9550 &8.0200 & 6.2878 &67.4\\
V11(9)  &16.561 &0.4318 & 0.0629 & 4.0100 &7.9779 & 5.4818 &3.8\\

\hline
\end{tabular}
\end{center}
n: number of harmonics used to fit the light curve.
}
\end{table*}

 The Fourier fit parameters have been used to derive
 physical parameters of the RR Lyrae stars.
For the RRc stars we employ the calibrations of Simon \& Clement (1993) and Morgan et al. (2007) and for the RRab stars the calibrations of Jurcsik \& Kov\'acs (1996),  Kov\'acs \& Jurcsik (1996) and Jurcsik (1998). The results are reported in {Tables  ~\ref{fisicosC} and {\ref{fisicosAB}, after taking into account the transformations to appropriate scales, as it will be discussed in the following subsections.

\subsection{[Fe/H]}

For the RRc stars, we have used the calibration of Morgan et al. (2007);

$$ {\rm [Fe/H]}_{ZW} = 52.466~P^2 ~-~ 30.075~P ~+~ 0.131~\phi^{(c)~2}_{31}  $$
\begin{equation}
~~~~~~~	~-~ 0.982 ~ \phi^{(c)}_{31} ~-~ 4.198~\phi^{(c)}_{31}~P ~+~ 2.424,
\end{equation}

\noindent
where $\phi^{(c)}_{31}$ is the phase in a series of cosines, and P the period in days. This calibration provides iron abundances in the metallicity scale of Zinn \& West (1984) with a standard deviation of 0.14 dex. The values of [Fe/H]$_{ZW}$ for the RRc stars are listed in column 2 of Table ~\ref{fisicosC}.

For the RRab stars, the calibration of Jurcsik \& Kov\'acs (1996) has been employed;

\begin{equation}
	{\rm [Fe/H]}_{J} = -5.038 ~-~ 5.394~P ~+~ 1.345~\phi^{(s)}_{31},
\end{equation}

\noindent
where $\phi^{(s)}_{31}$ is the phase in a sine series and before we used it, we transformed
our $\phi^{(c)}_{31}$ into $\phi^{(s)}_{31}$. The phases in sine and cosine series are correlated as :
  ~~ $\phi^{(s)}_{jk} = \phi^{(c)}_{jk} - (j - k) {\pi \over 2}$. The standard deviation of this calibration is 0.14 dex (Jurcsik 1998). 

The metallicity scale of eq. 5 differs from the ZW scale but they are related as ${\rm[Fe/H]}_{J} = 1.431 {\rm[Fe/H]}_{ZW} + 0.88$ (Jurcsik 1995). Thus, the difference between these two metallicity scales depends on the value of [Fe/H]; they coincide for [Fe/H]$\sim -2.0$ while for [Fe/H]$\sim -1.5$,
the [Fe/H]$_{J}$ is about 0.24 dex less metal-poor than [Fe/H]$_{ZW}$ (see also Fig. 2 of 
Jurcsik 1995). Therefore, for a metal poor cluster such as NGC 5466, the two scales are not significantly different. 

Eq. 5 is applicable to RRab stars with a  {\it deviation parameter} $D_m$, defined by
Jurcsik \& Kov\'acs (1996) and Kov\'acs \& Kanbur (1998), not exceeding a proper limit.
These authors suggest $D_m \leq 3.0$. In order to reinforce the statistical weight of the mean physical parameters we included stars with $D_m \leq 4.0$. $D_m$ values for individual stars are given in Table ~\ref{foufit}.
The metallicities [Fe/H]$_{J}$ and [Fe/H]$_{ZW}$ for the RRab stars
are listed IN COLUMNS 2 AND 3 OF Table ~\ref{fisicosAB}. We derive a mean values [Fe/H]$_{J}$= $-1.71 \pm 0.18$ and [Fe/H]$_{ZW}$= $-1.81 \pm 0.12$. It has been found however, from comparison with spectroscopic values of giant stars in clusters, that
eq. 5, at the low-metallicity end, predicts higher metallicities by $\sim$~0.2 dex (Kov\'acs 2002). Also a good agreement between the Fourier  and the spectroscopic [Fe/H] values for the LMC RR Lyrae stars was found by Gratton et al. (2004) and Di Fabrizio et al. (2005) after this metallicity scale difference is taken into account. Thus, following these authors, by subtracting 0.2 dex to the average [Fe/H]$_{J}$ we conclude that the mean iron abundance for the RRab stars in the ZW scale is $-1.91 \pm 0.18$. This value coincides within the uncertainties  with the mean [Fe/H]$_{ZW}$=1.81 $\pm 0.12$ calculated by the transformation proposed by Jurcsic (1995). 

Given the uncertainties in the above calibrations and the intrinsic scatter in the individual determinations of [Fe/H]$_{ZW}$, the corresponding mean values for RRc and RRab stars, $-1.92 \pm 0.21$ and $-1.91 \pm 0.18$ respectively, agree quite well.

For the RR Lyrae stars in the LMC it has been found that, while the metallicities derived from the Fourier decomposition may be discrepant from the spectroscopic determinations on a star-to-star comparison, the two approaches produce consistent mean values once the differences between metallicity scales are properly taken into account (Gratton et al. 2004). 
Therefore, it seems reasonable to conclude that the average iron content of NGC~5466, as derived from the RRab and RRc stars in the ZW metallicity scale, is [Fe/H]=$-1.91 \pm 0.19$. 

This value can be compared to the numerous recent iron content determinations available in the literature; [Fe/H] =$-1.6 \pm 0.3$ from intermediate resolution spectroscopy of giant stars (Pilachowski et al. 1983), [Fe/H] =$-2.17 \pm 0.12$ from UBV photometry of integrated light (Zinn 1980), [Fe/H] =$-1.87 \pm 0.11$ from a similar technique (Bica \& Pastoriza 1983), [Fe/H] =$-2.22$ from photographic photometry of complete samples of stars (Buonanno et al. 1985). It is generally accepted that NGC~5466 is one of the clusters with lowest metallicity content (e.g. Harris 1996; Salaris \& Weiss 2002) with values between $-2.2$ and $-2.1$. The iron content as estimated from the Fourier decomposition approach is relatively larger than these values but it is in
good agreement with larger estimates found in the literature.

\subsection{The luminosity log $L/L_{\odot}$ or $M_V$}

The luminosity or absolute magnitude for the RRc stars can be calculated using the theoretical calibration of Simon \& Clement (1993)

\begin{equation}
	log~L/L_\odot = 1.04~log~P - 0.058~\phi^{(c)}_{31} ~+~ 2.41. 
\end{equation}

As in previous papers (Arellano Ferro et al. 2004; 2006, L\'azaro et al. 2006) we note that
the luminosities of the RRc stars derived from eq. 6 are too large and inconsistent with the position of the theoretical instability strip boundaries (Bono et al. 1995) and HB evolutionary models (Lee \& Demarque 1990). This discrepancy has also been commented by Cacciari et al. (2005) in terms of the $M_V$-[Fe/H] relationship.

An alternative is the empirical calibration of Kov\'acs (1998)

\begin{equation}
M_V(K) = 1.261 ~-~ 0.961~P ~-~ 0.044~\phi^{(s)}_{21} ~-~ 4.447~A_4,  	
\end{equation}

\noindent
with an error of 0.042 mag. The luminosity scale of this calibration is based on the Baade-Wesselink luminosity scale.
Cacciari et al. (2005) have argued that the zero point of eq. 7 should be decreased by 0.2$\pm$0.02 mag in order to bring the absolute magnitudes into agreement with the mean magnitude for the RR Lyrae stars in the LMC, $V_0 = 19.064 \pm 0.064$ (Clementini et al. 2003). Following Cacciari et al. (2005) we report individual absolute magnitudes of the RRc stars in Table~\ref{fisicosC} after correcting the zero point of eq 7. to 1.061. The mean value is $M_V=0.53 \pm 0.06$. This values is therefore in agreement with a distance modulus of the LMC of $18.5 \pm 0.1$ (Freedman et al. 2001; van den Marel et al. 2002; Clementini et al. 2003).

For the RRab stars we have used the calibration of Kov\'acs \& Walker (2001) 

\begin{equation}
M_V(K) = ~-1.876~log~P ~-1.158~A_1 ~+0.821~A_3 +K,
\end{equation}

\noindent
which has an standard deviation of 0.04 mag.

The zero point of eq. 8, K=0.43 has been calculated by Kinman (2002) using the star RR Lyrae as calibrator adopting for RR Lyrae the absolute magnitude $M_V= 0.61 \pm 0.10$ mag, as derived by Benedict et al.  (2002) using the star parallax measured by the HST. 
Kinman (2002) finds his result to be consistent with the coefficients of the $M_V$-[Fe/H] relationship given by Chaboyer (1999) and Cacciari (2003). All these results are consistent with the distance modulus of the LMC of $18.5 \pm 0.1$.

We used eq. 8 to calculate individual values of $M_V$ for the RRab stars. We have included the two Blazhko stars V4 and V11 at maximum Blazhko amplitude since it has been shown that at such phase they behave like normal RRab stars (Jurcsik et al. 2002; Cacciari et al. 2005). The results are shown in Table ~\ref{fisicosAB}. The average is $0.52 \pm 0.11$. This result is in the same scale and in good agreement with the mean absolute magnitude found for the RRc stars.

\begin{table*}
\footnotesize{
\begin{center}
\caption[Parametros estelares de las RR Lyrae del tipo  c] {\small  Physical parameters for the 
RR{\lowercase {c}} stars }
\label{fisicosC}
\hspace{0.01cm}
 \begin{tabular}{lccccc}
\hline 
Star& [Fe/H]$_{ZW}$ & $M_V(K)$ & log$(L/L_{\odot})$ & $T_{\rm eff}$ &$D$ (kpc)\\
\hline
V12 & -1.581 & 0.63 & 1.658 & 7366.  & 15.44 \\
V13 & -2.062 & 0.53 & 1.697 & 7172.  & 16.59 \\
V15 & -1.847 & 0.48 & 1.719 & 7174.  & 16.54 \\
V16 & -2.008 & 0.48 & 1.716 & 7096.  & 15.42 \\
V17$^\ast$ & -2.040 & 0.57 & 1.677 & 6990.  & 14.23 \\
V18 & -2.085 & 0.55 & 1.689 & 7149. & 16.07 \\
V20$^\ast$ & -1.818 & 0.66  & 1.663 & 7300. & 15.27 \\
\hline
average & -1.92& 0.53 & 1.696 & 7191.4 &  16.01 \\
$\sigma$ & $\pm$0.21 & $\pm$0.06&$\pm$0.025 & $\pm$102.5 &  $\pm$0.56 \\
\hline
\end{tabular}
\end{center}
$\ast$: stars not included in the averages.
}
\end{table*}

\begin{table*}
\footnotesize{
\begin{center}
\caption[Parametros estelares de las RR Lyrae del tipo ab] {\small Physical parameters for the RR{\lowercase {ab}}
  stars } 
\label{fisicosAB}
\hspace{0.01cm}
 \begin{tabular}{lcccccc}
\hline 
Star&[Fe/H]$_{J}$ &[Fe/H]$_{ZW}$ & $M_V(K)$ & log$(L/L_{\odot})$ & $T_{\rm eff}$  &$D$ (kpc)\\
\hline
 V3        &-1.736 &-1.828  & 0.56 & 1.693 & 6405. & 15.97 \\
 V4$^{Bl}$ &-1.971 &-1.993  & 0.60 & 1.676 & 6472. & 16.36 \\
 V6        &-1.597 &-1.731  & 0.53 & 1.707 & 6340. & 16.38 \\
 V7        &-1.387 &-1.584  & 0.48 & 1.732 & 6253. & 15.75 \\
 V8        &-1.843 &-1.903  & 0.69 & 1.714 & 6297. & 16.03 \\
 V9        &-1.688 &-1.795  & 0.52 & 1.646 & 6312. & 14.50 \\
 V10$^*$   &-1.505 &-1.667  & 0.39 & 1.771 & 6246. & 17.28 \\
 V11$^{Bl}$&-1.811 &-1.881  & 0.41 & 1.752 & 6417. & 16.98 \\
 V14       &-1.678 &-1.788  & 0.36 & 1.785 & 6128. & 16.08 \\
\hline
average & -1.71&-1.81 & 0.52 & 1.720 & 6328. & 16.01\\
$\sigma$ &$\pm$0.18 &$\pm$0.12 & $\pm$0.11 & $\pm$0.045 & $\pm$108. & $\pm$0.71 \\
\hline
\end{tabular}
\end{center}
$Bl$: Blazhko stars with $D_m \leq 4.0$ at maximum amplitude, included in the average.\\
$\ast$: $D_m \geq 4.0$ not included in the average.
}
\end{table*}
  
The values of $M_V(K)$ in Tables~\ref{fisicosC} and \ref{fisicosAB} were transformed into $log~L/L_\odot$. The bolometric corrections for the average temperatures of RRc and RRab stars, given in Tables~\ref{fisicosC} and \ref{fisicosAB}, were estimated from the 
$T_{\rm eff}$-BC$_V$ from the models of Castelli (1999) as tabulated in Table 4 of Cacciari et al. (2005).
We adopted the value $M_{bol}^{\odot} = 4.75.$ These luminosities are plotted on Fig.{\ref{HRD}} and show a rather large dispersion. However, the mean value of both RRc AND RRab STARS, $log~L/L_\odot$=1.71, is in agreement with the predicted value by the theoretical models
included in the figure. This corroborates the result of Cacciari et al. (2005) that the Fourier coefficients can be used to estimate average luminosities of groups of stars while the individual values may not be trusted.

\subsection{The effective temperature $T_{\rm eff}$}

For the RRc stars we have used the calibration of Simon \& Clement (1993);

\begin{equation}
	log T_{\rm eff} = 3.7746 ~-~ 0.1452~log~P ~+~ 0.0056~\phi^{(c)}_{31}.
\end{equation}

For the RRab stars we used the calibrations of Jurcsik (1998)

\begin{equation}
	log~T_{\rm eff}= 3.9291 ~-~ 0.1112~(V - K)_o ~-~ 0.0032~[Fe/H],
\end{equation}

\noindent
with 

$$ (V - K)_o= 1.585 ~+~ 1.257~P ~-~ 0.273~A_1 ~-~ 0.234~\phi^{(s)}_{31} ~+~ $$
\begin{equation}
~~~~~~~ ~+~ 0.062~\phi^{(s)}_{41}.
\end{equation}

\noindent
Eq.10 has a standard deviation of 0.0018 (Jurcsik 1998), but the accuracy of $log~T_{\rm eff}$ is mostly set by the colour eq. 11. The error estimate on  $log~T_{\rm eff}$ is 0.003 (Jurcsik 1998).

It has been noticed by Cacciari et al. (2005) that the temperatures computed from the above equations do not match the colour-temperature relations predicted by the temperature scales of Sekiguchi \& Fukugita (2000) (SF) or by the  evolutionary models of Castelli (1999). Cacciari et al. (2005) compare the Fourier temperatures with ($B-V$) temperatures from the calibration of SF for RRc and RRab stars in M3 (see their Fig. 17). Clear trends and differences in temperature as large as 500~K are shown. We have repeated this comparison for the ($V-K$) temperatures of Cacciari et al. and similar trends and differences are found.
We have used these trends in an attempt to bring the temperature scales in eqs. 9, 10 and 11 onto the scale of SF or Castelli (1999). 

The temperatures from eqs. 9, 10 and 11
range from 6128~K to 6472~K for RRab stars and from 7000~K to 7366~K for RRc stars.
For the range of temperatures of RRab stars the agreement with SF scale is pretty good and no significant correction is necessary. The results are listed in Table~\ref{fisicosAB}. For the RRc stars the situation is more complicated.
The temperature difference and the temperature in the SF scales are correlated by the expression  $\Delta T_{\rm eff} (FourierVK-SF) = -(0.6793 \pm 0.0643) T_{SF} + (5032.9 \pm 474.9)$. For the range of temperatures of RRc stars, the temperature differences range from about 30~K at the hot end to 280~K at the cool end.
Fourier temperatures listed in   
Table~\ref{fisicosC}, in principle, should be reduced by the above differences to bring them into the SF temperature scale of the RR Lyraes in M3 (Cacciari et al. 2005).

Fig.{\ref{HRD} shows the positions of the RRc and RRab stars on the HR diagram. The luminosities and temperatures are those in Tables~\ref{fisicosC} and ~\ref{fisicosAB}. As a reference, in Fig.\ref{HRD} there are shown two models of the Zero Age Horizontal Branch (ZAHB) from VandenBerg, Bergbusch \& Dowler (2006) for [Fe/H]=$-1.836$ and 
[Fe/H]=$-2.012$ to allow for the uncertainties in the [Fe/H] determination. 
The empirical instability strip for RRab stars of Jurcsik (1998) is also shown after we
apply to it the luminosity correction discussed in section 4.2. 

The distribution on Fig.{\ref{HRD} of both the RRc and the RRab stars is clumpy. The RRc stars are too blue while the RRab stars are distributed in a very narrow vertical band. This narrow "instability strip" for the RRab stars was determined by Jurcsik (1998) using the Fourier  decomposition technique for 272 stars but it cannot be reconciled with the theoretical instability strip for the fundamental mode of Bono et al. (1995).
the large gap between both groups of stars is contrary to observational evidences in clusters with larger populations of RR Lyrae stars, like M3, where the distribution across the instability strip is more even and some fundamental and first overtone pulsator stars share  the inter-mode region (Cacciari et al. 2005).

In Fig.{\ref{HRD} we have also plotted the RRc stars with the temperatures brought to the SF temperature scale as described above (crosses). While the SF temperatures position the RRc stars closer to the RRab, closing the gap between the two groups, they fall much to the red of the first overtone red edge of the instability strip of Bono et al. (1995). 

The narrow distribution in colours or temperatures calculated with eqs. 10 and 11 has been noticed and discussed by Cacciari et al. (2005) and Di Fabrizio et al. (2005). The distribution of Fourier temperatures displayed in Fig. 12 confirms that result, and reinforces the conclusion of these authors that the temperatures obtained from the present Fourier parameters calibrations are uncertain. 
 
\begin{figure} 
\begin{center}
\includegraphics[width=8.5cm,height=7.5cm]{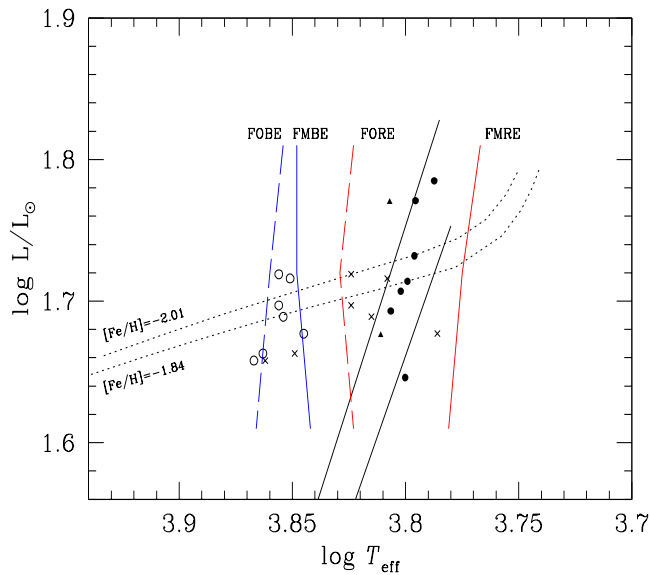}
\caption{Dots represent RRab stars, triangles Blazhko RRab stars and open circles RRc stars. $log(L/L_\odot)$ values were calculated from $M_V(K)$ in Tables \ref{fisicosC} and \ref{fisicosAB}.
The solid tilted lines indicate the empirical bounds of the RRab stars band found by Jurcsik (1998) 
from 272 stars after the zero point correction. The vertical boundaries are the fundamental 
(continuous lines) and first overtone (dashed lines) instability strips from 
Bono et al. (1995) for 0.65 $M/M_\odot$.
Two models of the ZAHB (VandenBerg, Bergbusch \& Dowler 2006) are shown (dashed lines) for [Fe/H]=$-1.836$ and $-2.012$.}
    \label{HRD}
\end{center}
\end{figure}

\subsection{The mass $M/M_{\odot}$}

The masses of the RRc stars can be estimated by the calibration of Simon \& Clement (1993):

\begin{equation}
log~M/M_\odot = 0.52~log~P ~-~ 0.11~\phi^{(c)}_{31} ~+~ 0.39,
\end{equation}

\noindent
and for the RRab stars by the calibration of Jurcsik (1998);

$$ log~M/M_\odot =20.884 ~-~ 1.754~log~P ~+~ 1.477~log~L/L_{\odot} $$

\begin{equation}
~~~~~ ~-~ 6.272~log~T_{\rm eff} ~+~ 0.367[Fe/H].
\end{equation}

However, Cacciari et al. (2005) have shown that the masses obtained for RR Lyrae stars from the above formulations and the Fourier decomposition of their light curves
are not reliable. Therefore we refrain from calculating the masses in the present paper.

\section{Discussion}

\subsection{Distance to NGC 5466}

To calculate the distance to the cluster we have calculated the distance moduli $(A_0 - M_V)$ 
 for each star, using the absolute magnitudes $M_V(K)$, given in {Tables  ~\ref{fisicosC} and {\ref{fisicosAB}, and the mean magnitudes $A_0$ from {Table ~\ref{foufit}}.
The value of $E(B - V)= 0.0$ was adopted (Harris 1996).
 
We find a mean distance of $16.01 \pm 0.56$ kpc for the RRc and
$16.01 \pm 0.71$ kpc for the RRab stars respectively. 
The uncertainty in these values is the standard deviation of the mean.
The mean distance to NGC~5466 from thirteen RRab and RRc variables is then of $16.0 \pm 0.6$ kpc. This  distance is in agreement with the mean luminosity of the RR Lyrae stars in the LMC $V_o=19.064 \pm 0.064$ (Clementini et al. 2003) and a distance modulus of $18.5 \pm 0.1$ mag for the LMC (Freedman et al. 2001; van den Marel et al. 2002; Clementini et al. 2003). It can be compared with the previous estimates found in the literature: 20.5 kpc from $PV$ photometry (Cuffey 1961); 16.3 kpc from $UBV$ integrated light
(Zinn 1980); 15.9 kpc from the HB level and its dependence on [Fe/H] (Harris 1996). Our value is therefore in good agreement with most recent photometric determinations.

\subsection{The $M_V$-[Fe/H] relationship}

This relationship has been amply discussed in the literature in the last decade from empirical standpoints (e.g. Lee et al. 1990; Salaris et al. 1997; Caloi et al 1997; Fernley et al. 1998; Lee \& Carney 1999; Chaboyer 1999; Demarque et al. 2000; Clementini et al. 2003; Cacciari \& Clementini 2003; Gratton et al. 2003, 2004; Rich et al. 2005) and theoretical grounds (Cassisi et al. 1999; Ferraro et al. 1999; Caputo et al. 2000; VandenBerg et al. 2000).  We felt it was worthwhile exploring this relation using the parameters derived from the Fourier decomposition of RR Lyrae
 light curves in globular clusters. There are about a dozen clusters in the literature, whose mean physical parameters have been calculated with the Fourier decomposition technique. A summary of those results, collected from the literature, can be found in Tables 7 and 8 of L\'azaro et al. (2006). The metallicities for the RRc stars are not on a homogeneous scale, and we opted for recalculating [Fe/H]$_{ZW}$ with the calibration of Morgan et al. (2007) only for those clusters for which we have access to the Fourier coefficients of the individual light curves, i.e. those clusters published by our group; 
M2, NGC~4147, M15 and NGC~5466 (L\'azaro et al. 2006; Arellano Ferro et al. 2004, 2006; present work, respectively). For the RRab stars 
the metallicities are all in the Jurcsik (1995) scale. The absolute magnitudes in those tables were calculated with the Kov\'acs (1998) calibration for the RRc stars and with the calibration of Kov\'acs \& Jurcsik (1996) for the RRab stars. 
We have recalculated the corresponding values for the cluster NGC~6229 studied by Borissova et al. (2001). This cluster was inadvertently  omitted by L\'azaro et al. (2006). 
Therefore, before discussing the $M_V$-[Fe/H] relationship we have made an effort to convert the metallicities and absolute magnitudes to homogeneous scales. Following the discussion in sections 4.1 and 4.2, we converted all metallicities to the ZW scale and the absolute magnitudes to be consistent with the luminosities  of the RR Lyraes in the LMC and with its distance modulus of 
$18.5 \pm 0.1$ mag.  
The list of clusters and their values of [Fe/H]$_{ZW}$ and $M_V(K)$ are listed in
Table~\ref{MVFEH}.

\begin{table}
\footnotesize{
\begin{center}
\caption[Valores medios para cumulos] {Mean [Fe/H]$_{ZW}$ and $M_V(K)$ derived from the RR Lyrae light curve Fourier decomposition for a family of clusters. Values taken from the compilation of L\'azaro et al. (2006) and converted to homogeneous scales as described in sections 4.1 and 4.2.}
\label{MVFEH}
\hspace{0.01cm}
 \begin{tabular}{lccc}
\hline 
Cluster& Oo & [Fe/H]$_{ZW}$& $M_V(K)$ \\
\hline
From RRab stars&&\\
\hline
NGC~4147 & I & $-1.42$ & 0.60 \\
NGC~6171 & I & $-1.21$ & 0.65 \\  
NGC~1851 & I & $-1.42$ & 0.60 \\   
M5      & I & $-1.43$ & 0.61 \\ 
M3      & I & $-1.62$ & 0.58 \\ 
NGC~6934 & I & $-1.73$ & 0.61 \\ 
NGC~6229 & I & $-1.60$ & 0.61 \\
NGC~5466 &II & $-1.91$ & 0.52 \\
M55     &II & $-1.68$ & 0.51 \\  
M2      &II & $-1.67$ & 0.51 \\
M92     &II & $-2.07$ & 0.47 \\
M15     &II & $-2.07$ & 0.47 \\
\hline
From RRc stars&&\\
\hline
NGC~4147 & I & $-1.53$ & 0.57 \\ 
NGC~5466 &II & $-1.92$ & 0.52 \\
M15     &II & $-2.09$ & 0.52 \\
M2      &II & $-1.80$ & 0.51 \\
\hline

\end{tabular}
\end{center}

}
\end{table}

Although some suggestions of a non-linear form of the $M_V$-[Fe/H] relationship exist from theoretical grounds (Dorman 1992; Cassisi et al. 1999; Caputo et al. 2000; Rey et al. 2000; VandenBerg et al. 2000), most empirical studies and some theoretical works support the relationship of the form $M_V= \alpha\rm{[Fe/H]} + \beta$. The values of $\alpha$ and $\beta$ found in the literature range from 0.17 to 0.30 and from 0.82 to 0.93 respectively, although there seems to be a growing consensus for a shallow slope $\alpha \sim 0.2$ and zero point  $\beta = 0.6$ for [Fe/H]$= -1.5$ (Cacciari \& Clementini 2003; Gratton et al. 2004).

Fig.~{\ref{FEHMV} shows the distribution of globular clusters with the values given in  Table \ref{MVFEH}. The solid line has the form: $
M_V=+(0{\rm .}18\pm0.03) \rm{[Fe/H]}+(0.85\pm0.05) $, that can be written as:

\begin{eqnarray}
M_V=+(0{\rm .}18\pm0{\rm .}03)([Fe/H]+1.5)+(0{\rm .}58\pm0{\rm .}05).
\end{eqnarray}

The above equation predicts the position of the RR Lyrae HB as a function of iron content and defines the RR Lyrae stars distance scale. According to the zero point corrections for the RR Lyrae [Fe/H] and $M_V$ values, discussed in sections 4.1 and 4.2, this equation is set in agreement with the distance modulus $18.5 \pm 0.1$ and metallicity [Fe/H]$=-1.5$ for the LMC. 
It predicts $M_V=0.58$ mag for  [Fe/H]=$-1.5$, in excellent agreement with the average absolute magnitude for RR Lyrae stars $M_V=0.59\pm 0.03$mag estimated by Cacciari \& Clementini (2003) from several independent methods. The correlation does not show evidence of non-linearity.
Eq. 15 can also be compared with very numerous theoretically predicted relationships whose coefficients
($\alpha$,$\beta$) are:
(0.17,0.56) (Lee et al. 1990); (0.21,0.59) (Salaris et al. 1997); (0.26,0.52)  (Caloi et al. 1997) and with empirically derived relationships with coefficients: (0.22,0.49) (Gratton et al. 1997); (0.20,0.68) (Fernley et al. 1998); (0.18,0.63) (Carretta et al. 2000); (0.214,0.56) (Clementini et al. 2003); (0.22,0.56) (Gratton et al. 2003).

\begin{figure} 
\includegraphics[width=8.5cm,height=6.cm]{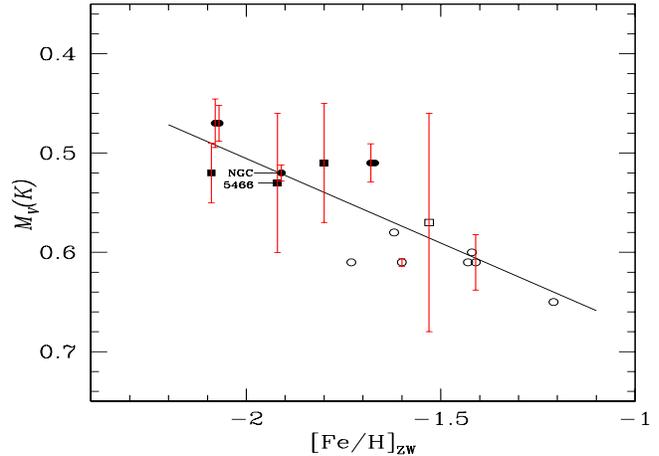}
\caption{The $M_V$-[Fe/H] relationship from the RR Lyrae Fourier light curve decomposition for a family of globular clusters. Metallicities and absolute magnitudes have been  transformed to the scales discussed in the text. Circles represent the results from the RRab stars while squares are from the RRc stars. Open symbols are used for OoI type clusters and filled symbols for OoII type clusters. NGC 5466 from the present work is labeled. The fitting line has the form $M_V=+(0{\rm .}18\pm0{\rm .}03)[Fe/H]+(0{\rm .}85\pm0{\rm .}05)$.}
    \label{FEHMV}
\end{figure}

It must be stressed at this point that, in order to obtain eq. 15, we first set the values of $M_V$ and [Fe/H] in specific broadely accepted scales and therefore, our derivation is not by itself an independent determination of the $M_V$-[Fe/H] relationship. However, the agreement of eq. 15 with other empirical and theoretical
versions of the $M_V$-[Fe/H] relationship shows that 
the Fourier decomposition of RR Lyrae stars light curves, if properly set to adequate scales, can be used to estimate reliable distances and metalicities to stellar systems.

\subsection{On the age of NGC~5466}

Ages of globular clusters have been estimated from the so called "vertical" and "horizontal" methods, i.e. estimating $\Delta$$V$, the difference of magnitudes of the ZAHB and the Turn Off (TO) point and $\Delta$$(V-I)$ or $\Delta$$(B-V)$, the difference between the TO and a fiducial point on the RGB (Rosenberg et al. 1999; Salaris \& Weiss 2002).  Both $\Delta$$V$ and $\Delta$$(V-I)$ are age and metallicity dependent.
These quantities for NGC~5466, 3.35$\pm$0.09 and 0.300$\pm$0.009 respectively, were measured by Rosenberg et al. (1999) from their homogeneous sample of globular cluster colour magnitude diagrams.
Adopting [Fe/H]$_{CG}=-2.13$, in the Carretta \& Gratton (1997) metallicity scale, and by confronting the $\Delta$$V$ and  $\Delta$$(V-I)$ parameters with different sets of isochrones, Rosenberg et al. (1999) calculated an age of 13.0$\pm$0.8 Gyr while Salaris \& Weiss (2002), using a different set of isochrones found 12.2$\pm$0.9 Gyr for NGC~5466. If the metallicity scale of ZW is used instead, [Fe/H]$_{ZW}=-2.22$, the age of NGC 5466 turned out to be 12.5$\pm$0.9 Gyr (Salaris \& Weiss 2002).

The calculation of theoretical values of $\Delta$$V$ requires the adoption of a relation between the HB absolute magnitude and the metal content. Rosenberg et al. (1999) adopted 
$M_V(ZAHB) = 0.18$[Fe/H] +0.92 (taken from Carretta et al. 2000). The alternative use of eq. 15 and the use of our value of [Fe/H]=$-1.91$, do not have a significant effect on the cluster age determination since they would produce a ZAHB only 0.03 mag dimmer, thus the cluster would appear slightly younger but well within the uncertainties of the determinations by Rosenberg et al. (1999) and Salaris \& Weiss (2002).

\section{Conclusions}

The technique of difference imaging has proven to be a powerful tool in order to find new variables even in crowded image regions. Here we report the discovery of five red semiregular variables, and the independent detection of three known eclipsing binaries and of faint SX Phe stars.

Physical parameters of astrophysical relevance, $log~(L/L_{\odot})$, 
 $log~T_{\rm eff}$, $M_V$ and [Fe/H], have been derived for the RR Lyrae stars in NGC~5466
using the Fourier decomposition of their light curves. Special attention has been paid to the conversion of these parameters into broadly accepted scales.
It has been found that $T_{\rm eff}$ for the RRc stars from the calibration of Simon \& Clement (1993), and for RRab stars from the calibration of Jurcsik (1998), cannot be reconciled with the theoretical predictions of the instability strip bounds. Also, these temperatures produce
unexpected distributions of both RRc and RRab stars on the HB, confirming the conclusion of Cacciari et al. (2005) and Di Fabrizio et al. (2005) that the temperatures derived from the above calibrations of the Fourier parameters are uncertain.

Proper transformations are required to 
set the mean values of $M_V$ and [Fe/H] of the RR Lyraes in adequate scales. Once this is done, the parameters derived for a family of globular clusters define a  $M_V$-[Fe/H] relationship
consistent with the metallicity and distance scales determined most recently from the RR Lyraes in the LMC (Gratton et al. 2004; Clementini et al. 2003).
The values for NGC 5466 are [Fe/H]=$-1.91 \pm 0.18$ and $D=16.0 \pm 0.6$ kpc, or a true distance modulus of $16.02 \pm 0.09$.
The above results and the isochrones of Salarsis \& Weiss (2002) imply an age of NGC~5466 slightly younger, by about 0.8 Gyr, than previous estimates. 

\begin{table}
\footnotesize{
\begin{center}
\caption[Nuevas variables] {\small Standard $V$ magnitudes of RR Lyrae stars in NGC 5466 (portion).}
\label{tab8}
\hspace{0.01cm}
\begin{tabular}{ccc}
\hline
Star & HJD & $V$\\

\hline
V2&   2453104.2887&	17.614\\
V2&   2453104.3231&	17.664\\	
V2 &  2453104.3295&	17.655\\	
V2 &  2453104.3357&	17.661\\	
V2 &  2453104.3419&	17.657\\	
...&&\\
\hline
\end{tabular}
\end{center}
}
\end{table}

\begin{table}
\footnotesize{
\begin{center}
\caption[Nuevas variables] {\small Instrumental $r$ magnitudes of RR Lyrae stars in NGC 5466 (portion).}
\label{tab9}
\hspace{0.01cm}
\begin{tabular}{ccc}
\hline
Star & HJD & $r$\\

\hline
V2 &  2453104.3262&16.815\\
V2 &  2453104.3326&16.819\\
V2  & 2453104.3388&16.829\\
V2 &  2453104.3526&	16.812\\
V2 &  2453104.3657&	16.806\\
...&&\\
\hline
\end{tabular}
\end{center}
}
\end{table}

\begin{table}
\footnotesize{
\begin{center}
\caption[Nuevas variables] {\small Standard $V$ magnitudes of new variables in NGC 5466 (portion).}
\label{tab10}
\hspace{0.01cm}
\begin{tabular}{ccc}
\hline
Star & HJD & $V$\\

\hline
V37 & 2453104.3388&	15.730\\
V37&  2453104.3526&	15.732\\	
V37 & 2453104.3657&	15.730\\	
V37 & 2453104.3789&	15.737\\	
V37 & 2453104.3851&	15.738\\	
...&&\\
\hline
\end{tabular}
\end{center}
}
\end{table}

\begin{table}
\footnotesize{
\begin{center}
\caption[Nuevas variables] {\small Standard $V$ magnitudes of three known eclipsing bianries (V28, V29 and V30)
and six SX Phe (NH29, NH30, NH35, NH38, NH39, NH49) stars in NGC 5466 (portion).}
\label{tab11}
\hspace{0.01cm}
\begin{tabular}{ccc}
\hline
Star & HJD & $V$\\

\hline
V28 & 2453104.3975&	18.554\\	
V28 & 2453104.4037&	18.511\\	
V28 & 2453104.4098&	18.546\\	
V28 & 2453104.4160&	18.540\\	
V28 & 2453104.4400&	18.551\\	
...&&\\
\hline
\end{tabular}
\end{center}
}
\end{table}

\section*{Acknowledgments}

We are grateful to the support astronomers of IAO, at Hanle and CREST (Hosakote) and at VBO (Kavalur), for their very efficient help while acquiring the data. AAF and VRL acknowledge support from DGAPA-UNAM grant through project IN108106. DMB is
thankful to the Instituto de Astronom\1a of the Universidad Nacional Aut\'onoma de M\'exico for their hospitality and to the Mexican Academy of Sciences and to the Royal Astronomical Society of London for financial support. We are grateful to Dr. E. Poretti for fruitful discussions. Numerous corrections, comments and suggestions by the referee, Dr. Gisella Clementini, contributed substantially to the improvement of the paper for which we are indebted. This work has made a large use of the SIMBAD and ADS services, for which we are thankful.

\end{document}